\documentclass[fleqn,10pt]{wlscirep}
\usepackage[utf8]{inputenc}
\usepackage[T1]{fontenc}
\usepackage{bbold}

\newcommand{\be}{\begin{equation}}
\newcommand{\ee}{\end{equation}}
\newcommand{\bea}{\begin{eqnarray}}
\newcommand{\eea}{\end{eqnarray}}
\newcommand{\nn}{\nonumber}
\newcommand{\eps}{\varepsilon}
\newcommand{\vk}{{\vec k}}

\newcommand{\bq}{{\boldsymbol{q}}}
\newcommand{\bk}{{\boldsymbol{k}}}
\newcommand{\bc}{{\boldsymbol{c}}}
\newcommand{\bd}{{\boldsymbol{d}}}

\newcommand{\br}{{\boldsymbol{r}}}

\newcommand{\D}{\Delta}
\newcommand{\up}{\uparrow}
\newcommand{\down}{\downarrow}
\newcommand{\lc}{\langle}
\newcommand{\rc}{\rangle}

\title{Synthetic complex Weyl superconductors, chiral Josephson effect and synthetic half-vortices}

\author[1,*]{Zahra Faraei}
\author[2,3]{Seyed Akbar Jafari}
\affil[1]{Institute for Advanced Studies in Basic Sciences (IASBS), Department of Physics, Zanjan, 45137-66731, Iran}
\affil[2]{Sharif University of  Technology, Department of Physics, Tehran, 11155-9161, Iran}
\affil[3]{Institute for Advanced Simulations, Forschungszentrum J\"ulich, 52425 J\"ulich, Germany}

\affil[*]{z.faraei@iasbs.ac.ir}

\affil[+]{these authors contributed equally to this work}

\begin{abstract}
We show that the most generic form of spin-singlet superconducting order parameter for chiral fermions is of the $\Delta_s+i\gamma^5\Delta_5$ where $\Delta_s$ is the usual
order parameter and $\Delta_5$ is the pseudo-scalar order parameter. After factoring out the $U(1)$ phase $e^{i\phi}$, this form of superconductivity admits yet additional complex 
structure in the plane of $(\Delta_s,\Delta_5)$. The polar angle $\chi$ in this plane dubbed chiral angle will be locked to the $U(1)$ phase $\phi$.  
We propose a synthetic setup based on stacking of topological insulators (TIs) and superconductors (SCs). Alternatively flux biasing the superconductors with a 
fluxes $\pm\Phi$ leads to $\Delta_5=\Delta_0 \sin(\chi)$, where $\Delta_0$ is the superconducting order parameter of the SC layers, and the chiral angle 
$\chi=\Phi/\Phi_0$ is directly given by the flux $\Phi$ in units of the flux quantum $\Phi_0=h/(2e)$. 
This can be used as a building block to construct a two-dimensional Josephson array. In this setup $\chi$ will be a background field defining a pseudoscalar $\Delta_5$
that can be tuned to desired configuration. While in a uniform background field $\Delta_5$ the dynamics of $\phi$ is given by standard XY model and its associated vortices, 
a {\em staggered} background $\pm\Delta_5$ (or equivalently $\chi$ and $\chi+\pi$ in alternating lattice sites) creates a new set of minima for the $\phi$ field that
will support half-vortex excitations. An isolated single synthetic "half-vortex" in the $\chi$ field in an otherwise uniform background will bind a $\phi$-half-vortex. 
This is similar to the way a p-wave superconducting vortex core binds a Majorana fermion.
\end{abstract}
\begin{document}

\flushbottom
\maketitle

\thispagestyle{empty}

\section*{Introduction}
 Weyl semimetals (WSMs) are a novel class of topological materials that host {\em chiral} fermions as their 
low-energy excitations~\cite{Yan2017,Hasan2015,Hasan2021}. The chirality is an additional attribute of the electrons in WSMs that defines whether the momentum and the spin
are parallel $(\tau=+1)$  or anti-parallel $(\tau=-1)$~\cite{Zeebook,Hasan2021}.
Having this extra attribute, the electrons in the WSMs will be described by four components. 
Extension of the phenomenological Ginzburg-Landau theory to a relativistic formalism~\cite{Capelle1999_1,Capelle1999_2,Ohsaku2001,Ohsaku2002} reveals different forms of superconductivity in these materials~\cite{WeylSup,Cho2012,bednik2015,Faraei2017}. Strictly speaking, the quartic fermionic expressions which represent the superconducting interactions can be made from  scalar, pseudoscalar, vector, axial vector and tensor bilinear  
structures under Lorentz transformations (for more details refer to supplementary material).
Practically, placing a three dimensional Dirac/Weyl semimetal in proximity to a conventional superconductor (with spin-singlet s-wave order parameter) can lead to all of these types of superconductivity due to the superconducting potential which is penetrated into the Dirac/Weyl matter~\cite{Faraei2017,Fu2010,Fu2014}. 

In this paper, we consider a model for Weyl superconductors that in addition to the conventional pairing (denoted by $\Delta_s$, with "s" subscript for scalar), 
simultaneously supports a novel form of pseudoscalar superconductivity denoted by $\Delta_5$ for obvious reasons. 
This model further allows one to adjust the values of the above two forms of order parameter by external flux bias. 
Our model builds on the model of Meng and Balents~\cite{WeylSup} according to which  a periodic stacking of 
magnetically doped topological insulator (TI) and a conventional superconductor gives rise to a Weyl superconductor~\cite{WeylSup}. 
In their work, all the superconducting layers have the same phase. We extend their model by allowing the superconducting phase of the
superconducting layers to alternate between $\chi$ and $-\chi$ (within a single building block). 
Tuning the phase $\chi$ allows us to realize not only the pure scalar~\cite{WeylSup} and pure pseudoscalar~\cite{Bovenzi2017} superconductivity,
but also a more interesting combination of them. Synthetic model of a combined form of superconductivity will be the main ingredient of our work. 
We will show that in a Josephson array composed of such superconductors, the interplay of these two kinds of superconducting orders 
which are respectively even and odd under parity, leads to synthesis of half-vortices. 
We show that the above phase variable $\chi$ (controlled by external bias) is given by the ratio of the pseudoscalar to the scalar component of the 
general superconducting order parameter and indeed is the polar angle in the complex plane of $(\Delta_s,\Delta_5)$. This {\em chiral angle} (CA)
plays a significant role in the Josephson energy and the Josephson current. 
The spatial variations of $\chi$ in a single Josephson junction
lead to a chiral Josephson current which in some circumstances can be separated from ordinary non-chiral one.

 This paper is structured as follows: First we introduce a synthetic setup for a tunable realization of a combination of scalar and pseudoscalar 
superconductivity. Our model is based on a generalization of the Meng and Balents model\cite{WeylSup} of Weyl superconductors. 
 It is followed by a discussion of the properties of such a combination of scalar and pseudoscalar superconductivity.
 Then we study a single Josephson junction of this type of synthetic superconductors and show the non trivial dependence of Josephson energy to the CA difference of the two superconductors. In the next section we discuss the synthesis of half-vortices in an array of this type of Weyl superconductors and then we drive the chiral Josephson current  in such arrays. The last section is devoted to summary and discussion.

\begin{figure}[t]
\centering
\includegraphics[width=10cm]{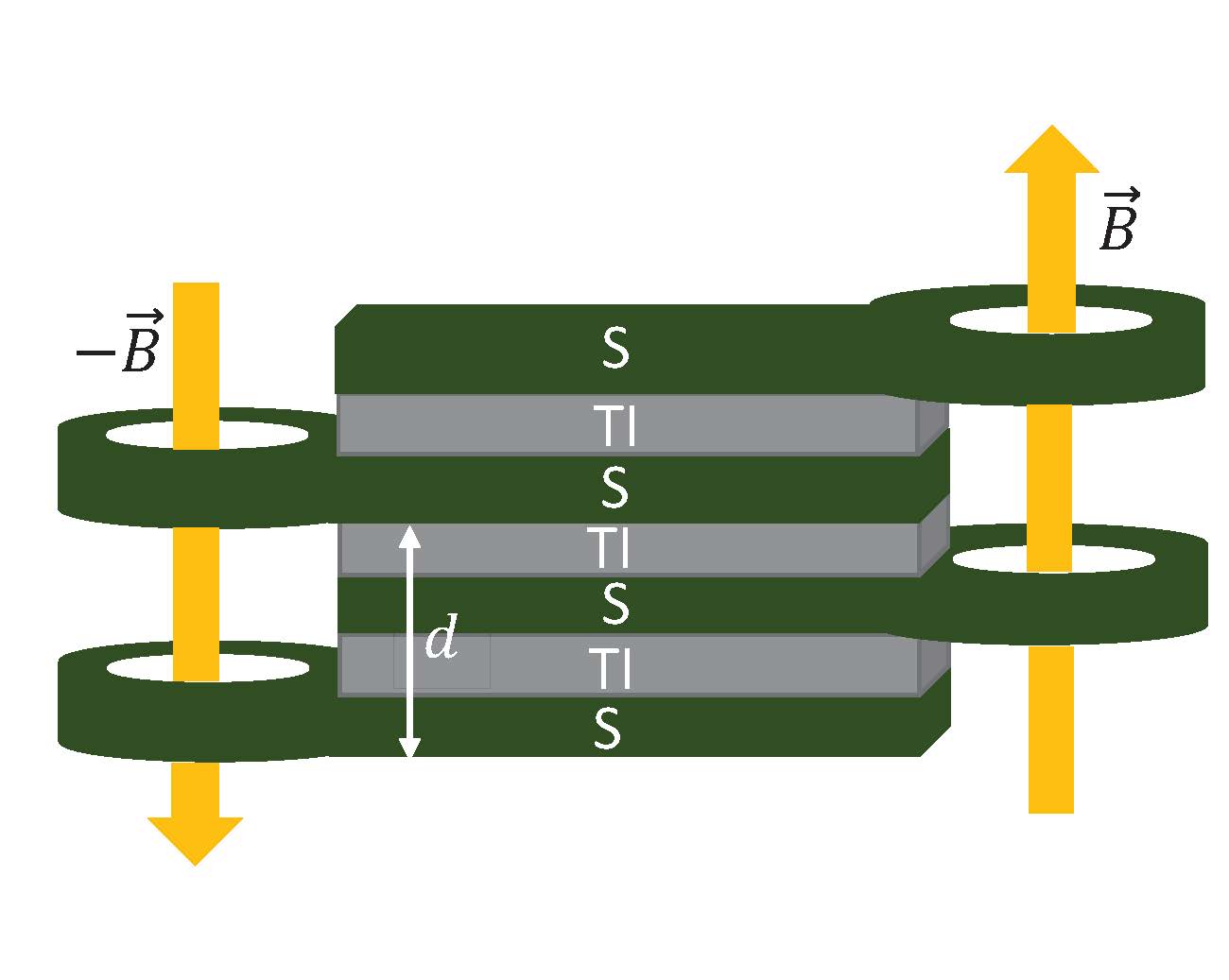}
\caption{A schematic model of a synthetic $(\Delta_5,\Delta_s)$ superconductor realized by stacking superconductor (SC) and topological insulator (TI) layers with a periodicity of $d$. A magnetic field $\vec{B}$ induces a flux $\Phi$ through the superconductors, which modifies the phases of adjacent SC layers by $\phi \pm \Phi/\Phi_0$, where $\phi$ is the initial uniform phase of the superconductors and $\Phi_0$ is the flux quantum.}
\label{s-model.fig}
\end{figure}

\section*{Results}

\subsection*{A model for synthetic realization of $(\Delta_s,\Delta_5)$ superconductor}
As Meng and Balents proposed in Ref.~\citeonline{WeylSup},
alternating stack of (s-wave) superconductors(SCs) and topological insulators (TIs) realizes a  Weyl superconductor. 
In this periodic structure, the tunneling parameter between the upper and lower layers of  each TI ($t_S$) differs from the tunneling parameter of a TI layer to the next TI layer ($t_D$). This consideration in addition to a magnetization ($m$) in the direction perpendicular to the surface of the layers, lead to the formation of Weyl nodes at $\vk=(0,0,\pi/d\pm k_0)$ where $k_0=1/d~\arccos (1-\frac{m^2-(t_S - t_D)^2}{2~t_S~t_D})$ and $d$ is the period of the structure. 
The role of superconductors is to proximitize the coupling potential and split each Weyl node to two Bogoliubov Weyl nodes. In their model, all superconductors have the same s-wave order parameter, $\Delta_0 e^{i\phi}$. Competition between $\Delta_0$ and $m$ results in four Weyl nodes at $\vk_=(0,0,\pi/d\pm k'_0)$ where $k'_0=1/d~\arccos (1-\frac{(m\pm\Delta_0)^2-(t_S - t_D)^2}{2~t_S~t_D})$.  In this way, the BdG Hamiltonian of this structure in basis $\psi=(\psi_{t\up},\psi_{t\down},\psi_{b\up},\psi_{b\down},\psi^*_{t\down},-\psi^*_{t\up},\psi^*_{b\down},-\psi^*_{b\up})$ where the subscripts $t$ and $b$  refer to the top and bottom surfaces of a TI layer and $\up\down$ refers to spin, is obtained as follows\cite{Bovenzi2017}.
\bea
\nn
H_\bk =v_f \eta_z \tau_z (k_y \sigma_x- k_x \sigma_y) + m\eta_0\tau_0\sigma_z+\eta_z(m_k \tau_x -\tau_y t_z \sin k_z d )\sigma_0 +\boldsymbol\Delta.
\eea
Pauli matrices 
$\bf\sigma$, $\bf\tau$ and $\bf\eta$ act in spin, top-bottom surfaces of TI layers and Nambu spaces, respectively and $m_k= t_s + t_D \cos (k_z d)$. The $8\times8$ matrix $\boldsymbol{\Delta}$ depend on the phase relationship of the top and bottom superconductors in each unit cell which consists of a TI layer and two SC layers surrounded it. If both superconductors have the same phase, $\boldsymbol{\Delta}=\Delta_0\eta_x \tau_0\sigma_0$ which is diagonal in 
 $\sigma$ and $\tau$  spaces and therefore the synthesised Weyl superconductor has the scalar pairing~\cite{WeylSup}. But, If top-bottom superconductors have a $\pi$ phase difference, $\boldsymbol{\Delta}=\Delta_0\eta_x \tau_z\sigma_0$, which changes sign upon inversion and is pseudoscalar under Lorentz transformation~\cite{Bovenzi2017}.
 
 Now, imagine setting the alternating phases to be $\Delta_0 e^{i(\phi+\chi)}$ and $\Delta_0 e^{i(\phi-\chi)}$ we obtain $\boldsymbol{\Delta}=\Delta_0 e^{i\phi}\eta_x (\cos \chi \tau_0 +i\sin\chi \tau_z)\sigma_0$. This naturally combines both scalar and pseudo-scalar superconducting orders and is a way to 
 synthesize a proximitized realization of a Weyl superconductor with the most general s-wave order parameter $\Delta_s+i\Delta_5 \gamma^5$ where $\Delta_s=\Delta_0 e^{i\phi} \cos\chi$ and $\Delta_5=\Delta_0 e^{i\phi} \sin\chi$. $\gamma^5$ in this expression is identified 
 as $\tau_z \sigma_0$ which is the same as $\gamma^5$ in Weyl representation. 
 
 As pointed out in the original paper~\cite{WeylSup}, the superconductors in the alternating stack of SC/TI can be flux biased in a way to induce desired phase 
 profile for the superconductors. This construction can be employed to design a situation where the phases of superconductors alternate between $\phi+\chi$ and $\phi-\chi$. 
From above construction, one produces a synthetic $\Delta_5$ that satisfies $\D_5/\D_s = \tan{\chi}$ where $\chi$ can be {\em tuned} by 
external flux bias as follows: Fig.~\ref{s-model.fig} shows a schematic representation of our synthetic superconducting system, which consists of a multilayer structure of SC and TI layers. The superconductors are subject to a magnetic field ($\vec B$) that induces alternating fluxes $\pm \Phi/\Phi_0$ ($\Phi_0$ is the flux quantum) on the SCs surrounding the TI layer. 
Therefore $\chi$ can be identified as $\Phi/\Phi_0$.
Raising $\chi$ from zero to $2\pi$ which corresponds to raising $\Phi$ from zero to $2\pi \Phi_0$, covers the entire angular span of the complex plane $(\D_s,\D_5)$. At $\chi=0$ where all the superconductors have the same phase, we have a conventional Weyl superconductor 
corresponding to $(\Delta_s=\Delta_0,\Delta_5=0)$ in the complex plane of $(\Delta_s,\Delta_5)$. 
At $\chi=\pi/2$ we have pure pseudoscalar superconductivity $(\Delta_s=0,\Delta_5=\Delta_0)$.
At $\chi=\pi$ we have $(\Delta_s=-\Delta_0,\Delta_5=0)$, and so on and so forth. 
Between the above limits when $\chi$ has a generic value, we will have a generic 
s-wave Weyl superconductor with order parameter $\D_s+i\D_5\gamma^5$. We note that an overall phase $e^{i\phi}$ does not affect our arguments. In other words, $\Delta_0$ could be complex and then factorizing the common phase factor of $\Delta_s$  and $\Delta_5$, the order parameter will be $e^{i\phi} (\D_s+i\D_5\gamma^5) $.

\begin{figure}[t]
\centering
\includegraphics[width=8cm]{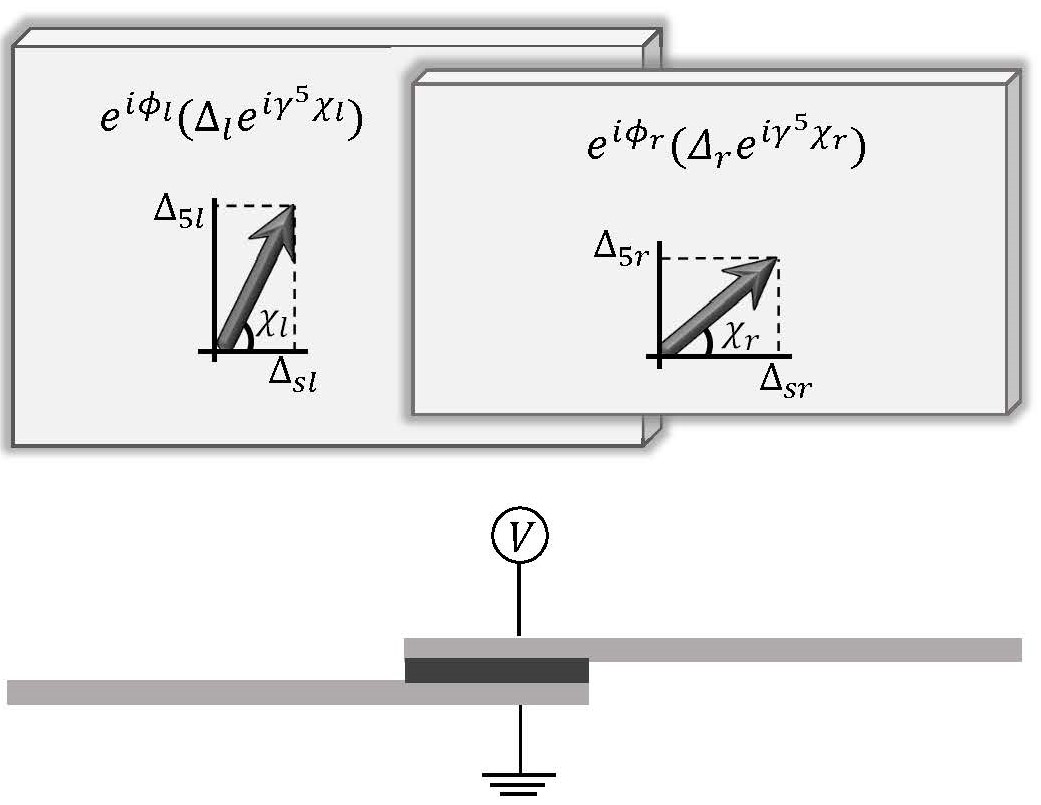}
\caption{(Color online) Top panel shows schematic top-view of the S$|$N$|$S junction. The lower panel
shows the side view. Each of $a=r/l$ (left/right) superconductors, the superconductivity is specified by
an strength $\Delta$ and {\em two phase angles} $(\phi,\chi)$. The Argand diagram in an emergent complex plane 
of $(\D_s,\D_5)$ defines the chiral angle $\chi$ as a new degree of freedom in superconducting Weyl semimetals }
\label{schematic.fig}
\end{figure}

\subsection*{Properties of the $(\Delta_s,\Delta_5)$ superconductivity}
The general pairing potential $e^{i\phi}(\D_s+i\D_5\gamma^5)$, which we showed 
can exist in Weyl superconductors manifestly breaks the standard (charge) $U(1)$ symmetry. But
under chiral gauge transformation $\Psi_e\to e^{i\gamma^5\theta/2}\Psi_e$ 
where the $\gamma^5$ is $\rm{diag}(1,1,-1,-1)$~\cite{Zeebook,Ryder1996},
we have, $\bar\Psi_e \to \bar\Psi_e e^{i\gamma^5\theta/2}$ and 
$\Psi_h\to e^{-i\gamma^5\theta/2}\Psi_h$. Therefore the pairing preserves the axial $U(1)$ symmetry in the Fermion sector.

 The diagonal elements of $\gamma^5$ are the chirality
eigenvalues and for real $\D_s$ and $\D_5$, the prefactor $i$ guarantees the Hermiticity of the resulting
Weyl-Bogoliubov-De Gennes (WBdG) Hamiltonian. 
Therefore the amplitude $\D_5$ denotes nothing but the previously found {\em pseudoscalar} superconductivity~\cite{Faraei2017}. 
Such form of pseudoscalar superconductivity, spontaneously breaks not only the global $U(1)$ symmetry 
(that is broken by every SC once a non-zero $\D_s$ is picked up), but also the parity symmetry
that corresponds to realizing one of the eigenvalues of $\gamma^5$ and giving the eigenvalues $e^{i\phi}(\D_s\pm i\D_5)$. 
Here choosing either of $\pm$ sings corresponds to spontaneously breaking a $Z_2$ symmetry. Hence, after factoring out an overall $U(1)$ phase, the remaining $\D_s$ and $\D_5$ will be both real numbers. Now,
consider the geometric algebra~\cite{ChrisBook} constructed from scalar $\mathbb{1}$ and 
the {\em pseudoscalar} $\gamma^0\gamma^1\gamma^2\gamma^3$ ($=-i\gamma^5$) of the Clifford algebra:
$\D_s {\mathbb 1}-\D_5\gamma^0\gamma^1\gamma^2\gamma^3$. This means apart from the $U(1)$ phase $e^{i\phi}$, the most generic form of superconductivity in WSMs can be represented by
a number in the complex plane $(\D_s,\D_5)$. The conventional superconductors are confined to the real axis of
this plane, while the pure pseudoscalar superconductors~\cite{Faraei2017,Salehi2017} are confined to the imaginary axis of this plane. 

In order to appreciate the importance of the notion of the complex $(\D_s,\D_5)$ plane, let us consider a purely 
pseudoscalar superconductor~\cite{Salehi2017}, $(0,\D_5)$. 
The complex plane structure allows to immediately understand the topologically non-trivial structure of such a superconductor. 
The coefficient "i"$=e^{i\pi/2}$ in front of $\D_5$, directly enters the 
amplitude of Andreev reflection at a superconductor-normal interface. 
This amounts to an additional phase change of $\pi/2$ upon every Andreev reflection (for more details refer to supplementary material). 
Therefore in a superconductor-normal-superconductor (S$|$N$|$S) Josephson junction, a total phase of $\pi$ is accumulated at the two interfaces 
of the S$|$N$|$S junction. 
Such a $\pi$ phase corresponds to change in the number parity. Therefore in a closed loop geometry of a Josephson junction, 
the electron has to traverse the loop once again, giving rise to the $4\pi$ (two rounds) periodic Josephson effect. 
The $4\pi$-periodic Josephson effect is a hallmark of topological superconductivity and its associated Majorana modes~\cite{Salehi2017}.
This can be the most natural explanation for the observed $4\pi-$periodic Andreev bound states in such systems~\cite{4pi2018}.
So the Josephson physics on the real axis ($\D_s$) and imaginary axis ($\D_5$) are significantly different. 
Now we are going to explore the rest of this complex plane and show that it contains a remarkably rich physics
of half-vortices and confinement transition.

\subsection*{Josephson coupling}

Alternative way of expressing the complex algebra $(\D_s,\D_5)$ of superconducting Dirac/Weyl materials 
is to specify it by an strength $\Delta$ and two angles $(\phi,\chi)$. $\phi$ being the $U(1)$ phase 
couples to external EM fields, while $\chi$ is the polar angle in the complex plane $(\D_s,\D_5)$.
This way pairing equation can be represented as 
\be
    e^{i\phi}(\D_s\mathbb{1}+i\D_5\gamma^5)=e^{i\phi}\Delta e^{i\gamma^5\chi}. 
    \label{dpid.eqn}
\ee
where $\Delta=\sqrt{\D_s^2+\D_5^2}$ is the amplitude of superconducting order parameter and the chiral angle (CA) is defined by $\tan\chi=|\D_5/\D_s|$.

Therefore we need to extend the Josephson physics that involves a single phase $\phi$, to include now a pair of phase variables $(\phi,\chi)$. 
Given two superconductors $\Delta_a e^{i\phi_a} e^{i\chi_a}$ where $a=l,r$ corresponds to left/right  
superconductor as in Fig.~\ref{schematic.fig}, how will the CAs, $\chi_a$ modify the Josephson effect? 
The answer to this fundamental question will provide us with the effective Hamiltonian governing the dynamics of
phase fields $(\phi,\chi)$ in Josephson arrays. 
The chemical potential difference between the left ($l$) and right ($r$) superconductors is set by the voltage $V$ across the barrier. 
We assume that the barrier layer is sufficiently thin for electrons to tunnel through, and that the 
tunneling process can be regarded as a small perturbation. The tunneling current is given by~\cite{Kita2015}
\bea
\label{ITr.eqn}
I = \dfrac{e}{\hbar^2} \dfrac{1}{v_l v_r} \sum_{\bk,\bq} \int_{-\infty}^t  dt' ~ e^{0_+ t'}  Tr\bigg[\lc \hat{\bc}_\bk (t) \hat{\bc}_\bk^\dagger (t') \rc  \hat{T}_{\bk\bq}  \eta_z  \lc \hat{\bd}_\bq^* (t) \hat{\bd}_\bq^T(t') \rc^T  \hat{T}_{\bq\bk} - \lc \hat{\bc}_\bk^* (t') \hat{\bc}_\bk^T(t) \rc^T  \hat{T}_{\bk\bq}  \eta_z  \lc \hat{\bd}_\bq (t') \hat{\bd}_\bq^\dagger (t) \rc  \hat{T}_{\bq\bk}\bigg], 
\eea
where $v_l (v_r)$  is the volume of the $l$ ($r$) superconductor, $\hat{\bc}_\bk$ and $\hat{\bd}_\bq$  are the electron field operators that annihilate in
$l$ and $r$ superconductors at wave vectors $\bk$ and $\bq$, respectively and $\hat{T}_{\bk\bq}$ is the tunneling matrix element between them. 
We further assume that the tunneling is independent of spin and chirality and does not break the 
time-reversal symmetry, $\hat{T}_{\bk\bq}=\hat{T}^*_{-\bq,-\bk}$. Therefore the tunneling matrix in the chirality, spin and Nambu space is specified
by the unit matrices, $\tau_0$, $\sigma_0$ and $\eta_0$, respectively: $\hat{T}_{\bk\bq}=t \sigma_0 \tau_0 \eta_0$. 
$\bf\sigma$, $\bf\tau$ and $\bf\eta$ are Pauli matrices act in spin, chirality(pseudo-spin corresponds to top-bottom layer of periodic TI-SC multilayer\cite{Meng2018BH}) and Nambu space, respectively.

In situations where some form of boundary conditions~\cite{Faraei2018,Witten2016, Akhmerov2008, McCann_2004} impose a chirality reversal during the tunneling, 
one has to make a replacement $\tau_0\to\tau_z$, where $\tau_z$ is the third Pauli matrix in the chirality space. 
We find that the later form of tunneling process does not change our main result. Hence, in what follows we focus on the simplest case given above. 
The factor $e^{0_+t'}$, guarantees that the integrand vanishes for $t' \rightarrow -\infty$ and thereby ensures the convergence of the $t'$ integral. 

It is straightforward to calculate the trace terms in current formula \eqref{ITr.eqn} by evaluating the expectation values as functions of $V$. It has two parts. 
One is the single particle part which is not relevant to Josephson current. Here we only focus on the second term that describes the transport of the Cooper pairs which includes the following term,
\bea
Tr\big[\lc \hat{\bc}_\bk (t) \hat{\bc}_\bk^\dagger (t') \rc \hat{T}_{\bk\bq} \eta_z  \lc \hat{\bd}_\bq^* (t) \hat{\bd}_\bq^T(t') \rc^T  \hat{T}_{\bq\bk}\big]
\label{trace.eqn} \supset
e^{ieV(t+t')/\hbar } f(\eps_\bk,\eps_\bq,t) -  e^{- ieV(t+t')/\hbar} f^*(\eps_\bk,\eps_\bq,-t),
\eea
where $f(\eps_\bk,\eps_\bq,t)$ is defined as
\bea
\dfrac{\D_l \D_r}{\eps_\bk \eps_\bq}   n_-(\eps_\bk) n_-(\eps_\bq) e^{i( \phi_l - \phi_r)}   \cos(\chi_l - \chi_r). 
\label{f.eqn}
\eea
Here $\eps_{\bk}=\sqrt{k_x^2+k_y^2+k_z^2+\D_{l}^2}$ is the dispersion relation for the excitations of the left superconductor. Similarly the  $\eps_{\bq}$
which is obtained by $k\to q$ and $\Delta_l\to\Delta_r$ is the dispersion relation of the right superconductor. Furthermore, 
$n_\pm(\eps_{\bk/\bq}) = n(\eps_{\bk/\bq}) e^{i \eps_{\bk/\bq} t/ \hbar} \pm n(-\eps_{\bk/\bq}) e^{-i \eps_{\bk/\bq} t/\hbar}$ 
where $n(\eps_{\bk/\bq})=(e^{\beta\eps_{\bk/\bq}}+1)^{-1}$ is the Fermi-Dirac distribution function. 
There are some other terms in equation~\eqref{trace.eqn} which include $(k_x \sigma_x + k_y \sigma_y)$ and $(k_x\sigma_y - k_y\sigma_x)$
but these terms give null contributions upon integration over $k_x$ and $k_y$, so we drop them. 
As it is seen in $f$, the dependence on the CA $\chi$ is of the $\cos(\chi_l-\chi_r)$ form. 
This result is robust against variations in the boundary conditions (i.e. replacing $\tau_0\to\tau_z$ in the tunneling matrix).

\subsection*{Half-vortices}

Substitution of ~\eqref{trace.eqn} in \eqref{ITr.eqn}, the total Josephson current becomes,
\bea
 I = I_{s} \cos(\chi_l-\chi_r), \label{current.eqn}
 \eea
where $I_{s}$ is the standard form of the AC Josephson current with frequency $2eV/h$ that also appears in conventional 
superconductors,
\bea
I_{s}=I_{ss} \sin(\dfrac{2eVt}{\hbar}+\phi_l-\phi_r)+ I_{sc} \cos(\dfrac{2eVt}{\hbar}+\phi_l-\phi_r).
\label{Is.eqn}
\eea
Here  $I_{ss}$ and $I_{sc}$ are functions of $\Delta_l$ and $\Delta_r$ such that at zero voltage 
the $I_{sc}$ becomes zero  but $I_{ss}$ has a finite value and determines the critical current. 
For more details we refer the readers to the supplementary material. Let us now focus at $V=0$ where $I_{sc}$ becomes zero and the second 
term disappears. In the absence of deriving voltage, the $I_s$ solely arises from the phase difference $\phi_l-\phi_r$ of the superconductors:
\bea
   I_s=I_{ss} \sin(\phi_l-\phi_r)\cos(\chi_l-\chi_r).
\eea
Generalization of a single Josephson junction into an array of Josephson junctions on a lattice whose sites are
labeled by $i,j$, yields
\bea
\label{phichi.eqn}
   I_{ij}=I_{ss} [\sin(\phi_{ij}+\chi_{ij})+\sin(\phi_{ij}-\chi_{ij})]/2,
\eea
where $\phi_{ij}=\phi_i-\phi_j$ and $ \chi_{ij}=\chi_i-\chi_j$. The above two terms are indeed very suggestive: if one defines new phase fields $\varphi^\pm_{ij}=\phi_{ij}\pm\chi_{ij}$~\cite{Bertlmann2000},
it can be interpreted as an ordinary Josephson current for the right/left phase variables $\varphi^\pm$. In the continuum, 
the $\phi$ couples to EM gauge field via $\partial_\mu\phi\to\partial_\mu\phi-(2e)A$.
The above two terms in the current, when integrated with respect to the phase difference $\phi_{ij}$ to obtain the energy,
will produce $-\cos(\phi_{ij}+\chi_{ij})-\cos(\phi_{ij}-\chi_{ij})$. Therefore the classical energy of 
the Josephson array will be 
\be
   E[{\phi_i,\chi_i}]= - \lambda_J \sum_{\langle ij\rangle} \cos(\phi_{ij})\cos(\chi_{ij}).
   \label{extendJosephson.eqn}
\ee

\begin{figure}[t]
\centering
\includegraphics[width=15cm]{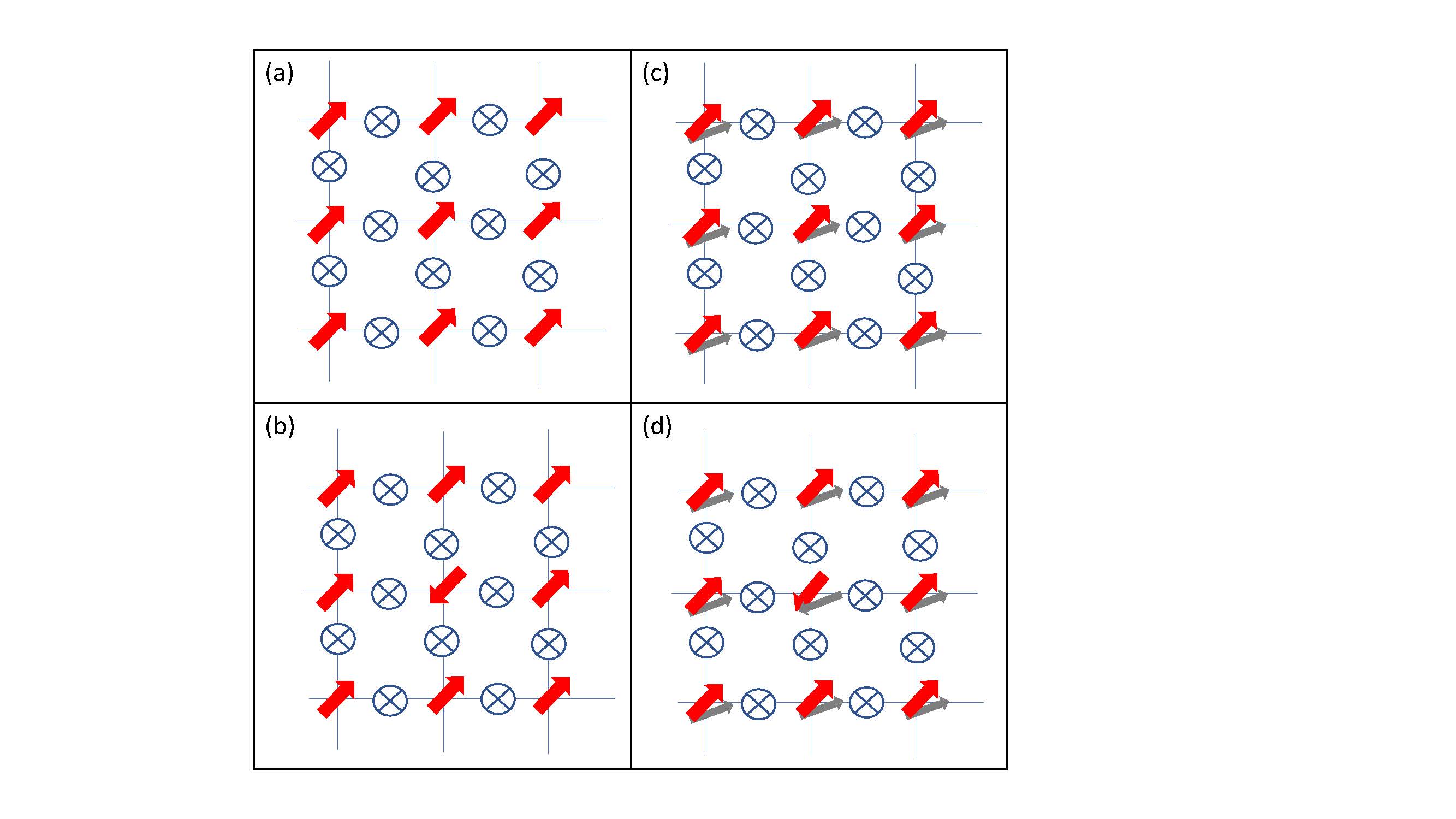}
\caption{(Color online) 
(a) and (b) represent the lowest energy state and an excited state of a 2D Josephson array of 
scalar superconductors, $\D_s$. Red arrows represent the phase variables $\phi_i$ and $\otimes$s are the Josephson junctions.
In (c) and (d) the chiral angles are also important and are denoted by gray arrows. 
One possible ground state for $\phi_i$ (red arrows). In this case, the CA phases also point
in their own fixed direction (d) The phase (red) of the second superconductor (counting from the left) is flipped. 
This effect can be compensated by an associated flip in the $\chi$. 
}
\label{arrows.fig}
\end{figure}

This equation is a generalization of the standard Josephson potential energy $-\sum_{\langle i,j\rangle}\cos{\phi_{ij}}$ 
(the classical XY model) which is minimized when all the superconductors of the Josephson array, have the same phase, namely $\phi_{ij}=0$. 
The ground state of the standard XY model is depicted by the red arrows in Fig.~\ref{arrows.fig} (a). 
Every out of phase junction has the energy cost $2\lambda_J$. For example the central superconductor in Fig.~\ref{arrows.fig} (b) is 
completely out of phase and costs $2\times 2\lambda_J$. 
For simplicity let us focus on the continuum limit of the Josephson array. 
At low temperatures, the partition function of classical XY model
is dominated by slowly varying configurations of the field $\phi(\br)$, and the $\cos\phi$ term will become $|\nabla\phi|^2$
to penalize  deviations from the uniform $U(1)$ phase configurations~\cite{Altland2009,Schakel2008}. The configurations in the excited states of
this model for which the identification $\phi\sim \phi\pm 2\pi$ round a circle
in the real space can be made, correspond to vortex/anti-vortex solutions. 

Now let us turn our attention to Fig.~\ref{arrows.fig} (c) and (d) where in addition to the $U(1)$ phase variables (red arrows),
the CA variables (gray arrows) also enter the game as a (pseudoscalar) background field. 
The important feature of equation~\eqref{extendJosephson.eqn} is a locking between the rotor variable $\phi$ of the XY model 
and background field $\chi$. 
While in the XY model, every phase (red) shift $\phi\to\phi+\pi$ such as the one 
in Fig.~\ref{arrows.fig}(b) entails an energy cost of $2\lambda_J$, in the extended XY model of equation~\eqref{extendJosephson.eqn}
this can be compensated by a corresponding flip in the chiral angle field (gray) $\chi$ of the two Weyl superconductors. 
While Fig.~\ref{arrows.fig}~(c) corresponds to the ground state in a uniform background $\chi$-field, 
by modifying the configurations of the background pseudoscalar $\chi$-field one may access a ground state such as Fig.~\ref{arrows.fig}~(d).
The same phase flip can be engineered for any other superconductor.
This situation bears certain similarity to half-vortices in superfluid $^3$He-A where the matrix of order parameter of superfluid is given by the 
product of its spin and orbital parts\cite{Mineev}, then a change of sign of orbital part of the order parameter acquired
over any closed path in the liquid corresponding to half quantum vortex can be compensated by the change of sign
of the spin part of the order parameter, so that the whole order parameter will be single-valued. But the present scenario differs
from the above example in that here the pseudoscalar field $\chi$ can be {\em externally tuned} to a synthetic half-vortex configuration.
Such a synthetic half-vortex of $\chi$ will bind a half-vortex of $\phi$ due to the locking in Eq.~\eqref{extendJosephson.eqn}.
This can be thought of trapping a $\phi$-half-vortex. The fact that topologically non-trivial configurations of the background field $\chi$
can bind a $\phi$-half-vortex is quite similar to the way in which the vortex core (topologically non-trivial configuration) of a p-wave 
superconductor binds Majorana fermions~\cite{Stone2004}. If there one synthesizes another $\chi$-half-vortex, it can bind the other $\phi$-half-vortex. Otherwise
the other $\phi$-half-vortex can only bind to the boundary of the system.

Eq.~\eqref{extendJosephson.eqn} is a novel classical statistical mechanics problem that deserves investigation of its own. Nevertheless a qualitative
physics of this model is evident from the following consideration. The Villain expansion of the ordinary XY model consists in representing the partition function
of a $\cos\theta$ interaction featuring minima at $2\pi n$ in terms of parabolas centered around these minima. Let us now think of a
completely different configuration of the $\chi$-field, namely a staggered configuration where it alternates its sign, one can generate new set of minima for $\cos\phi$ that are 
located at half-integer multiples of $2\pi$. In this way, the staggered pseudoscalar field $\chi$ gives rise to a set of parabolic minima located at
integer and half-integer multiples of $2\pi$. One can imagine a similar Villain expansion around all these minima. The simplest function that in addition to minima at integer
multiples of $2\pi$ also contains minima at half-integer multiples of $2\pi$ is achieved by an additional $\cos2\phi$ term. In this case the natural {\em excitations}
of the $\phi$-field will be half-vortices. On the other hand, uniform configuration of the background $\chi$ will favor full vortices that arise from the usual $\cos\phi$ term.
Therefore, a generic configuration of $\chi$ can be modelled by a competition between $\cos\phi$ and $\cos2\phi$~\cite{Chalker2017}.
Within this model, the superfluid phase of the extended XY model
as a condensate of the {\em dual boson pairs}~\cite{Chalker2017}, a generalization of the XY model which has been introduced in various fields and studied by
several authors (~\citeonline{GeneralizedXY} and references in it). 

The half vortices in a synthetic background $\chi$ field can be interpreted in terms of chirality imbalance in the following sense.
Imagine a 2D array of Josephson junctions. In the absence of synthetic $\D_5$, a configuration of red arrows that
round a closed path is identified as $\phi\sim\phi\pm2 \pi$ defines a vortex/anti-vortex (excited) state. 
As discussed above, when both $\D_s$ and $\D_5$ are present, the superconducting phase round a circle can also
be identified as $\phi\sim \phi\pm\pi$, which is accompanied by a corresponding synthesized (Ising) half-vortex in $\chi\sim\chi\mp\pi$.
Since a half-vortex in $\phi$ is always accompanied by a half-vortex in $\chi$, and that $\varphi_\pm=\phi\pm\chi$, the half-vortex state 
can be alternatively interpreted as chirality polarized vortex state where the population of vortices in $\varphi_+$ and $\varphi_-$ fields differs. 

Irrespective of our interpretation, the EM field only couples to $\phi$ and not to $\chi$.
This will then provide a very sharp and definite experimental signature for the half-vortex state as follows: Straightforwardly
following the arguments of Weinberg~\cite{WeinbergBook}, instead of obtaining the flux quantum $2\pi\hbar/(2e)=h/(2e)$ -- as in ordinary superconductors --
one finds the flux quantum $\pi\hbar/(2e)$. Interpreting this quantization rule as $h/(2\tilde e)$ as in conventional superconductors, one is
lead to conclude that $2\tilde e=4e$. This "doubling" can be immediately detected in the ac Josephson frequency of a single junction. 
Although it appears that we are dealing with a condensate of a {\em pair of Cooper pairs} (remember the boson doubling contained in $\phi\to 2\phi$),
but in fact the root cause of this effect is the presence of the CA $\chi$ that plays a compensating role in 
equation~\eqref{extendJosephson.eqn}. Such a frequency doubling can be regarded as manifestation of the staggered configuration of the $\chi$ field.

\subsection*{Chiral Josephson current}

So far, the chiral angle $\chi$ has lead to the formation of half-vortices and the associated confinement transition
that separates this phase from the full vortex state. 
In a similar way that spatial variations of $\phi$ in a conventional
superconductor leads to the conventional supercurrent, let us show that the spatial variations of $\chi$ leads to a
chiral Josephson current. According to TI/SC model, this situation could be realized by continuously varying the $\chi$ difference of neighbour Weyl superconductors of the array which is controlled by the phase difference of top and bottom superconductors of each cell.  To compute the chiral Josephson current, we return again to a single Josephson junction setup of Fig.~\ref{schematic.fig}. The only generalization we need
to perform in equation~\eqref{ITr.eqn} is to insert an additional tensor product of Pauli matrices $\tau_z\eta_z$. 
The $\tau_z$ encodes the fact that right- and left-handed chiral fermions have to enter the chiral current
with opposite signs. The $\eta_z$ encodes the fact that time reversal operation mapping the electrons and holes
in the BdG equation indeed flips the chirality. Therefore, 
\bea
I_{5} = \dfrac{e}{\hbar^2} \dfrac{1}{v_l v_r} \sum_{\bk,\bq} \int_{-\infty}^t  dt'~ e^{0_+ t'} Tr\bigg[ \lc \hat{\bc}_\bk (t) \hat{\bc}_\bk^\dagger (t') \rc  \hat{T}_{\bk\bq}  \eta_z(\tau_z\eta_z)  \lc \hat{\bd}_\bq^* (t) \hat{\bd}_\bq^T(t') \rc^T  \hat{T}_{\bq\bk} 
 - \lc \hat{\bc}_\bk^* (t') \hat{\bc}_\bk^T(t) \rc^T  \hat{T}_{\bk\bq}  \eta_z(\tau_z\eta_z)  \lc \hat{\bd}_\bq (t') \hat{\bd}_\bq^\dagger (t) \rc  \hat{T}_{\bq\bk} \bigg]. \nn
\eea
This equation yields the chiral Josephson current when the applied voltage $V=0$ as:
 \bea
  I_{5}
  =I_{ss}\sin{(\phi_l-\phi_r)}\sin(\chi_l-\chi_r). 
  \label{I5.eqn}
\eea
Note that the pair $(I,I_5)$ are both determined by the quantity $I_{ss}$ introduced in equation~\eqref{current.eqn}. 
The only difference is that, $I$ is proportional to $\cos(\chi_l-\chi_r)$, while $I_5$ is proportional to the $\sin(\chi_l-\chi_r)$. 
In both cases the phase difference $\phi_{lr}$ is needed to drive the current of Cooper pairs. When the CA difference $\chi_{lr}$ is non-zero, in addition to the electric current, a net chirality is also carried 
by the Cooper pairs. It is useful to view the pair of numbers $(I,I_5)$ in a complex plane 
and note that the modulus of this complex numbers will be $I_{ss}\sin(\phi_{lr})$ of non-chiral superconductors. 
Therefore, for a generic chiral angle difference $\chi_{lr}$ the Josephson current has both non-chiral ($I$) and chiral ($I_5$) "components". 
In the special case where the difference in the chiral angles of the two superconductors is $\chi_{lr}=\pi/2$, the non-chiral
Josephson current will be zero, and the entire Josephson current will become chiral (i.e. along the imaginary axis in the complex plane of $I$ and $I_5$). 

Taking the continuum limit of the above results immediately reveals that 
the spatial variations of the CA, $\chi$ generates chiral currents. 
In 1+1 dimensional spacetime the above chiral current acquires a nice interpretation as follows: 
The continuum limit of the Josephson lattice of superconductors of Weyl semimetals according to equation~\eqref{I5.eqn} will give 
$I_5\propto \chi(x+\delta x_a)-\chi(x)\propto \partial_a\chi$,
where $a$ denotes a spatial direction. Lorentz boosting this result gives, $I_{5,\mu}\propto \partial_\mu \chi$. 
In $1+1$ spacetime dimensions, using the fact that $\gamma^\mu=\varepsilon^{\mu\nu}\gamma_\nu\gamma^5$,
the above result immediately gives $I^\mu=\varepsilon^{\mu\nu}\partial_\nu\chi$ which has
a manifest Goldston-Wilczek current form~\cite{Schakel2008,Goldstone1981,Wilczek2002} and satisfies the conservation $\partial_\mu I^\mu=0$. 
This qualifies the chiral Josephson current in 1+1 dimension as a direct manifestation of Goldstone-Wilczek current.

\section*{Discussion}

The most generic form of a spin-singlet superconducting Weyl semimetal is specified by a single $U(1)$ phase $\phi$
and a pair of real numbers $(\D_s,\D_5)$ that form a complex algebra whose polar angle defines a chiral angle $\chi$. 
This angle does not result from spontaneous breaking of chiral $U(1)$ symmetry. It merely represents the polar angle in 
Argand diagram of an emergent complex plane $(\D_s,\D_5)$.
In this paper, we have shown how to synthesize such a $(\Delta_5,\Delta_s)$ superconductor by alternatively flux-biased
arrangements of BCS superconductors and topological insulators. As such the chiral angle $\chi$ can be directly controlled by
the amount of flux bias. Furthermore, a sign change of $\Delta_5\to -\Delta_5$ ($\chi\to\chi+\pi$) can be induced by 
replacing the flux bias in the left and right side of the building block shown in Fig.~\ref{s-model.fig}. In fact sign
reversal under reflection is a defining property of a {\em pseudo-}scalar. This setup will provide a synthetic framework
to {\em externally apply} a pseudoscalar background field $\chi$.
In fact application of pseudo-fields is an interesting topics of its own. In Ref.~\cite{Oka2016} it has been suggested that
the a circularly polarized light on Weyl semimetals acts as a pseudo-gauge field. In this respect, our model can be regarded
as much simpler setup where pseudo-gauge fields of finite strength can be externally applied to a system.

Depending on the configuration of the background pseudoscalar field $\chi$, a variety of interesting effects can be 
produced. (i) When one synthesizes an isolated half-vortex in the $\chi$ field, it will bind a $\phi$-half-vortex. This is 
similar to the way a vortex core in a p-wave superconductor binds a Majorana fermion~\cite{Stone2004}. In the case of an
isolated synthetic half vortex in the background $\chi$-field, the other $\phi$-half-vortex will be localized in the boundary of 
the system. (ii) For a staggered configuration of background $\chi$ field the additional minima generated by the alternating
sign of the $\cos\chi$ at half-integer multiples of $2\pi$ will re-arrange the ground state in a way that the $\phi$-half-vortices
will be supported. In a Josephson array based on our synthetic $(\Delta_s,\Delta_5)$ building element,
where the background field is neither uniform, not staggered, rather it could belong to a range of configurations in between, interesting 
competition between the tendency to support vortices versus tendency to support half-vortices will set in. 
A possible externally tuned phase transition between the full vortex and half vortex case will be a confinement-deconfinement phase
transition. The vortex Nernst-effect that continues to be present at temperatures above the BKT transition
has proven to be a reliable signature of vorticity~\cite{Ong2004}.

\section*{Methods}

We have used the algebra of Dirac matrices, Fierz identities and tunneling mechanism to calculate the Josephson current. The calculation methods are detailed in supplementary material.

\section*{Data Availability}

 The data supporting the findings of this study are available within the article and Supplementary material.

\bibliography{mybib}

\begin{thebibliography}{10}
\urlstyle{rm}
\expandafter\ifx\csname url\endcsname\relax
  \def\url#1{\texttt{#1}}\fi
\expandafter\ifx\csname urlprefix\endcsname\relax\def\urlprefix{URL }\fi
\expandafter\ifx\csname doiprefix\endcsname\relax\def\doiprefix{DOI: }\fi
\providecommand{\bibinfo}[2]{#2}
\providecommand{\eprint}[2][]{\url{#2}}

\bibitem{Yan2017}
\bibinfo{author}{Yan, B.} \& \bibinfo{author}{Felser, C.}
\newblock \bibinfo{journal}{\bibinfo{title}{Topological materials: Weyl
  semimetals}}.
\newblock {\emph{\JournalTitle{Annual Review of Condensed Matter Physics}}}
  \textbf{\bibinfo{volume}{8}}, \bibinfo{pages}{337--354},
  \doiprefix\url{10.1146/annurev-conmatphys-031016-025458}
  (\bibinfo{year}{2017}).

\bibitem{Hasan2015}
\bibinfo{author}{Xu, S.-Y.} \emph{et~al.}
\newblock \bibinfo{journal}{\bibinfo{title}{Discovery of a weyl fermion
  semimetal and topological fermi arcs}}.
\newblock {\emph{\JournalTitle{Phys. Rev. B}}} \textbf{\bibinfo{volume}{349}},
  \bibinfo{pages}{613--617}, \doiprefix\url{10.1126/science.aaa9297}
  (\bibinfo{year}{2015}).

\bibitem{Hasan2021}
\bibinfo{author}{Hasan, M.~Z.} \emph{et~al.}
\newblock \bibinfo{journal}{\bibinfo{title}{Weyl, dirac and high-fold
  chiral~fermions in topological quantum matter}}.
\newblock {\emph{\JournalTitle{Phys. Rev. B}}} \textbf{\bibinfo{volume}{6}},
  \bibinfo{pages}{784--803}, \doiprefix\url{10.1038/s41578-021-00301-3}
  (\bibinfo{year}{2021}).

\bibitem{Zeebook}
\bibinfo{author}{Zee, A.}
\newblock \emph{\bibinfo{title}{Quantum Field Theory in a Nutshell}}
  (\bibinfo{publisher}{Princeton University Press},
  \bibinfo{address}{Princeton}, \bibinfo{year}{2010}).

\bibitem{Capelle1999_1}
\bibinfo{author}{Capelle, K.} \& \bibinfo{author}{Gross, E. K.~U.}
\newblock \bibinfo{journal}{\bibinfo{title}{Relativistic framework for
  microscopic theories of superconductivity. i. the dirac equation for
  superconductors}}.
\newblock {\emph{\JournalTitle{Phys. Rev. B}}} \textbf{\bibinfo{volume}{59}},
  \bibinfo{pages}{7140--7154}, \doiprefix\url{10.1103/PhysRevB.59.7140}
  (\bibinfo{year}{1999}).

\bibitem{Capelle1999_2}
\bibinfo{author}{Capelle, K.} \& \bibinfo{author}{Gross, E. K.~U.}
\newblock \bibinfo{journal}{\bibinfo{title}{Relativistic framework for
  microscopic theories of superconductivity. ii. the pauli equation for
  superconductors}}.
\newblock {\emph{\JournalTitle{Phys. Rev. B}}} \textbf{\bibinfo{volume}{59}},
  \bibinfo{pages}{7155--7165}, \doiprefix\url{10.1103/PhysRevB.59.7155}
  (\bibinfo{year}{1999}).

\bibitem{Ohsaku2001}
\bibinfo{author}{Ohsaku, T.}
\newblock \bibinfo{journal}{\bibinfo{title}{Bcs and generalized bcs
  superconductivity in relativistic quantum field theory: Formulation}}.
\newblock {\emph{\JournalTitle{Phys. Rev. B}}} \textbf{\bibinfo{volume}{65}},
  \bibinfo{pages}{024512}, \doiprefix\url{10.1103/PhysRevB.65.024512}
  (\bibinfo{year}{2001}).

\bibitem{Ohsaku2002}
\bibinfo{author}{Ohsaku, T.}
\newblock \bibinfo{journal}{\bibinfo{title}{Bcs and generalized bcs
  superconductivity in relativistic quantum field theory. ii. numerical
  calculations}}.
\newblock {\emph{\JournalTitle{Phys. Rev. B}}} \textbf{\bibinfo{volume}{66}},
  \bibinfo{pages}{054518}, \doiprefix\url{10.1103/PhysRevB.66.054518}
  (\bibinfo{year}{2002}).

\bibitem{WeylSup}
\bibinfo{author}{Meng, T.} \& \bibinfo{author}{Balents, L.}
\newblock \bibinfo{journal}{\bibinfo{title}{Weyl superconductors}}.
\newblock {\emph{\JournalTitle{Phys. Rev. B}}} \textbf{\bibinfo{volume}{86}},
  \bibinfo{pages}{054504}, \doiprefix\url{10.1103/PhysRevB.86.054504}
  (\bibinfo{year}{2012}).

\bibitem{Cho2012}
\bibinfo{author}{Cho, G.~Y.}, \bibinfo{author}{Bardarson, J.~H.},
  \bibinfo{author}{Lu, Y.-M.} \& \bibinfo{author}{Moore, J.~E.}
\newblock \bibinfo{journal}{\bibinfo{title}{Superconductivity of doped weyl
  semimetals: Finite-momentum pairing and electronic analog of the
  ${}^{3}$he-$a$ phase}}.
\newblock {\emph{\JournalTitle{Phys. Rev. B}}} \textbf{\bibinfo{volume}{86}},
  \bibinfo{pages}{214514}, \doiprefix\url{10.1103/PhysRevB.86.214514}
  (\bibinfo{year}{2012}).

\bibitem{bednik2015}
\bibinfo{author}{Bednik, G.}, \bibinfo{author}{Zyuzin, A.~A.} \&
  \bibinfo{author}{Burkov, A.~A.}
\newblock \bibinfo{journal}{\bibinfo{title}{Superconductivity in weyl metals}}.
\newblock {\emph{\JournalTitle{Phys. Rev. B}}} \textbf{\bibinfo{volume}{92}},
  \bibinfo{pages}{035153}, \doiprefix\url{10.1103/PhysRevB.92.035153}
  (\bibinfo{year}{2015}).

\bibitem{Faraei2017}
\bibinfo{author}{Faraei, Z.} \& \bibinfo{author}{Jafari, S.~A.}
\newblock \bibinfo{journal}{\bibinfo{title}{Superconducting proximity in
  three-dimensional dirac materials: Odd-frequency, pseudoscalar, pseudovector,
  and tensor-valued superconducting orders}}.
\newblock {\emph{\JournalTitle{Phys. Rev. B}}} \textbf{\bibinfo{volume}{96}},
  \bibinfo{pages}{134516}, \doiprefix\url{10.1103/PhysRevB.96.134516}
  (\bibinfo{year}{2017}).

\bibitem{Fu2010}
\bibinfo{author}{Fu, L.} \& \bibinfo{author}{Berg, E.}
\newblock \bibinfo{journal}{\bibinfo{title}{Odd-parity topological
  superconductors: Theory and application to
  ${\mathrm{cu}}_{x}{\mathrm{bi}}_{2}{\mathrm{se}}_{3}$}}.
\newblock {\emph{\JournalTitle{Phys. Rev. Lett.}}}
  \textbf{\bibinfo{volume}{105}}, \bibinfo{pages}{097001},
  \doiprefix\url{10.1103/PhysRevLett.105.097001} (\bibinfo{year}{2010}).

\bibitem{Fu2014}
\bibinfo{author}{Fu, L.}
\newblock \bibinfo{journal}{\bibinfo{title}{Odd-parity topological
  superconductor with nematic order: Application to
  ${\mathrm{cu}}_{x}{\mathrm{bi}}_{2}{\mathrm{se}}_{3}$}}.
\newblock {\emph{\JournalTitle{Phys. Rev. B}}} \textbf{\bibinfo{volume}{90}},
  \bibinfo{pages}{100509}, \doiprefix\url{10.1103/PhysRevB.90.100509}
  (\bibinfo{year}{2014}).

\bibitem{Bovenzi2017}
\bibinfo{author}{Bovenzi, N.} \emph{et~al.}
\newblock \bibinfo{journal}{\bibinfo{title}{Chirality blockade of andreev
  reflection in a magnetic weyl semimetal}}.
\newblock {\emph{\JournalTitle{Phys. Rev. B}}} \textbf{\bibinfo{volume}{96}},
  \bibinfo{pages}{035437}, \doiprefix\url{10.1103/PhysRevB.96.035437}
  (\bibinfo{year}{2017}).

\bibitem{Ryder1996}
\bibinfo{author}{Ryder, L.~H.}
\newblock \emph{\bibinfo{title}{Quantum Field Theory}}
  (\bibinfo{publisher}{Cambridge University Press}, \bibinfo{year}{1996}).

\bibitem{ChrisBook}
\bibinfo{author}{Doran, C.} \& \bibinfo{author}{Lasenby, A.}
\newblock \emph{\bibinfo{title}{Geometric Algebra for Physicists}}
  (\bibinfo{publisher}{Cambridge University Press}, \bibinfo{year}{2003}).

\bibitem{Salehi2017}
\bibinfo{author}{Salehi, M.} \& \bibinfo{author}{Jafari, S.~A.}
\newblock \bibinfo{journal}{\bibinfo{title}{Sea of majorana fermions from
  pseudo-scalar superconducting order in three dimensional dirac materials}}.
\newblock {\emph{\JournalTitle{Scientific Reports}}}
  \textbf{\bibinfo{volume}{7}}, \doiprefix\url{10.1038/s41598-017-07298-2}
  (\bibinfo{year}{2017}).

\bibitem{4pi2018}
\bibinfo{author}{Li, C.} \emph{et~al.}
\newblock \bibinfo{journal}{\bibinfo{title}{4$\pi$-periodic andreev bound
  states in a dirac semimetal}}.
\newblock {\emph{\JournalTitle{Nature Materials}}}
  \textbf{\bibinfo{volume}{17}}, \bibinfo{pages}{875--880},
  \doiprefix\url{10.1038/s41563-018-0158-6} (\bibinfo{year}{2018}).

\bibitem{Kita2015}
\bibinfo{author}{Kita, T.}
\newblock \emph{\bibinfo{title}{Statistical Mechanics of Superconductivity}}
  (\bibinfo{publisher}{Springer Japan}, \bibinfo{year}{2015}).

\bibitem{Meng2018BH}
\bibinfo{author}{Liu, H.}, \bibinfo{author}{Sun, J.-T.},
  \bibinfo{author}{Huagn, H.}, \bibinfo{author}{Liu, F.} \&
  \bibinfo{author}{Meng, S.}
\newblock \bibinfo{journal}{\bibinfo{title}{Fermionic analogue of black hole
  radiation with a super high hawking temperature}}.
\newblock {\emph{\JournalTitle{arXiv:1809.00479}}}  (\bibinfo{year}{2018}).

\bibitem{Faraei2018}
\bibinfo{author}{Faraei, Z.}, \bibinfo{author}{Farajollahpour, T.} \&
  \bibinfo{author}{Jafari, S.~A.}
\newblock \bibinfo{journal}{\bibinfo{title}{Green's function of semi-infinite
  weyl semimetals}}.
\newblock {\emph{\JournalTitle{Physical Review B}}}
  \textbf{\bibinfo{volume}{98}}, \doiprefix\url{10.1103/physrevb.98.195402}
  (\bibinfo{year}{2018}).

\bibitem{Witten2016}
\bibinfo{author}{Witten, E.}
\newblock \bibinfo{journal}{\bibinfo{title}{Anomalies in quantum field
  theory}}.
\newblock {\emph{\JournalTitle{La Rivista del Nuovo Cimento}}}
  \textbf{\bibinfo{volume}{39}}, \bibinfo{pages}{313–370},
  \doiprefix\url{10.1393/ncr/i2016-10125-3} (\bibinfo{year}{2016}).

\bibitem{Akhmerov2008}
\bibinfo{author}{Akhmerov, A.~R.} \& \bibinfo{author}{Beenakker, C. W.~J.}
\newblock \bibinfo{journal}{\bibinfo{title}{Boundary conditions for dirac
  fermions on a terminated honeycomb lattice}}.
\newblock {\emph{\JournalTitle{Phys. Rev. B}}} \textbf{\bibinfo{volume}{77}},
  \bibinfo{pages}{085423}, \doiprefix\url{10.1103/PhysRevB.77.085423}
  (\bibinfo{year}{2008}).

\bibitem{McCann_2004}
\bibinfo{author}{McCann, E.} \& \bibinfo{author}{ko, V. I.~F.}
\newblock \bibinfo{journal}{\bibinfo{title}{Symmetry of boundary conditions of
  the dirac equation for electrons in carbon nanotubes}}.
\newblock {\emph{\JournalTitle{Journal of Physics: Condensed Matter}}}
  \textbf{\bibinfo{volume}{16}}, \bibinfo{pages}{2371--2379},
  \doiprefix\url{10.1088/0953-8984/16/13/016} (\bibinfo{year}{2004}).

\bibitem{Bertlmann2000}
\bibinfo{author}{Bertlmann, R.~A.}
\newblock \emph{\bibinfo{title}{Anomalies in Quantum Field Theory}}
  (\bibinfo{publisher}{Oxford University Press}, \bibinfo{year}{2000}).

\bibitem{Altland2009}
\bibinfo{author}{Altland, A.} \& \bibinfo{author}{Simons, B.~D.}
\newblock \emph{\bibinfo{title}{Condensed Matter Field Theory}}
  (\bibinfo{publisher}{Cambridge University Press}, \bibinfo{year}{2009}).

\bibitem{Schakel2008}
\bibinfo{author}{Schakel, A. M.~J.}
\newblock \emph{\bibinfo{title}{Boulevard of Broken Symmetries}}
  (\bibinfo{publisher}{{WORLD} {SCIENTIFIC}}, \bibinfo{year}{2008}).

\bibitem{Mineev}
\bibinfo{author}{Mineev, V.~P.}
\newblock \bibinfo{journal}{\bibinfo{title}{Half-quantum vortices}}.
\newblock {\emph{\JournalTitle{Low Temprature Physics}}}
  \textbf{\bibinfo{volume}{39}}, \bibinfo{pages}{818},
  \doiprefix\url{10.1063/1.4823487} (\bibinfo{year}{2013}).

\bibitem{Stone2004}
\bibinfo{author}{Stone, M.} \& \bibinfo{author}{Roy, R.}
\newblock \bibinfo{journal}{\bibinfo{title}{Edge modes, edge currents, and
  gauge invariance in ${p}_{x}{+ip}_{y}$ superfluids and superconductors}}.
\newblock {\emph{\JournalTitle{Phys. Rev. B}}} \textbf{\bibinfo{volume}{69}},
  \bibinfo{pages}{184511}, \doiprefix\url{10.1103/PhysRevB.69.184511}
  (\bibinfo{year}{2004}).

\bibitem{Chalker2017}
\bibinfo{author}{Serna, P.}, \bibinfo{author}{Chalker, J.~T.} \&
  \bibinfo{author}{Fendley, P.}
\newblock \bibinfo{journal}{\bibinfo{title}{Deconfinement transitions in a
  generalised {XY} model}}.
\newblock {\emph{\JournalTitle{Journal of Physics A: Mathematical and
  Theoretical}}} \textbf{\bibinfo{volume}{50}}, \bibinfo{pages}{424003},
  \doiprefix\url{10.1088/1751-8121/aa89a1} (\bibinfo{year}{2017}).

\bibitem{GeneralizedXY}
\bibinfo{author}{Canova, G.~A.}, \bibinfo{author}{Levin, Y.} \&
  \bibinfo{author}{Arenzon, J.~J.}
\newblock \bibinfo{journal}{\bibinfo{title}{Kosterlitz-thouless and potts
  transitions in a generalized $xy$ model}}.
\newblock {\emph{\JournalTitle{Phys. Rev. E}}} \textbf{\bibinfo{volume}{89}},
  \bibinfo{pages}{012126}, \doiprefix\url{10.1103/PhysRevE.89.012126}
  (\bibinfo{year}{2014}).

\bibitem{WeinbergBook}
\bibinfo{author}{Weinberg, S.}
\newblock \emph{\bibinfo{title}{The Quantum Theory of Fields, Vol. II}}
  (\bibinfo{publisher}{Cambridge University Press}, \bibinfo{address}{New
  Delhi}, \bibinfo{year}{2005}).

\bibitem{Goldstone1981}
\bibinfo{author}{Goldstone, J.} \& \bibinfo{author}{Wilczek, F.}
\newblock \bibinfo{journal}{\bibinfo{title}{Fractional quantum numbers on
  solitons}}.
\newblock {\emph{\JournalTitle{Physical Review Letters}}}
  \textbf{\bibinfo{volume}{47}}, \bibinfo{pages}{986--989},
  \doiprefix\url{10.1103/physrevlett.47.986} (\bibinfo{year}{1981}).

\bibitem{Wilczek2002}
\bibinfo{author}{Wilczek, F.}
\newblock \bibinfo{title}{{SOME} {BASIC} {ASPECTS} {OF} {FRACTIONAL} {QUANTUM}
  {NUMBERS}}.
\newblock In \emph{\bibinfo{booktitle}{World Scientific Series in 20th Century
  Physics}}, \bibinfo{pages}{135--152},
  \doiprefix\url{10.1142/9789812777041_0017} (\bibinfo{publisher}{{WORLD}
  {SCIENTIFIC}}, \bibinfo{year}{2002}).

\bibitem{Oka2016}
\bibinfo{author}{Ebihara, S.}, \bibinfo{author}{Fukushima, K.} \&
  \bibinfo{author}{Oka, T.}
\newblock \bibinfo{journal}{\bibinfo{title}{Chiral pumping effect induced by
  rotating electric fields}}.
\newblock {\emph{\JournalTitle{Phys. Rev. B}}} \textbf{\bibinfo{volume}{93}},
  \bibinfo{pages}{155107}, \doiprefix\url{10.1103/PhysRevB.93.155107}
  (\bibinfo{year}{2016}).

\bibitem{Ong2004}
\bibinfo{author}{Ong, N.}, \bibinfo{author}{Wang, Y.}, \bibinfo{author}{Ono,
  S.}, \bibinfo{author}{Ando, Y.} \& \bibinfo{author}{Uchida, S.}
\newblock \bibinfo{journal}{\bibinfo{title}{Vorticity and the nernst effect in
  cuprate superconductors}}.
\newblock {\emph{\JournalTitle{Annalen der Physik}}}
  \textbf{\bibinfo{volume}{13}}, \bibinfo{pages}{9--14},
  \doiprefix\url{10.1002/andp.200310034} (\bibinfo{year}{2004}).

\end{thebibliography}

\section*{Acknowledgements}

S. A. J. was supported by grant No. G960214 from the research deputy of Sharif University of Technology
and Iran Science Elites Federation (ISEF). The revision of this work was done while SAJ was supported
by Alexander von Humboldt fellowship.
We thank Abolhassan Vaezi for fruitful discussions. 

\section*{Author contributions statement}

Z. F. performed the mathematical calculations and provided the figures. The manuscript was jointly written by both authors. Both authors discussed and commented on the results and reviewed the manuscript.

\section*{Additional information}

The authors declare no competing financial interests.

\end{document}